\documentclass[11pt,english]{article}
\usepackage[T1]{fontenc}
\usepackage[latin9]{inputenc}
\usepackage{geometry}
\usepackage{color}
\usepackage{array}
\usepackage{multirow}
\usepackage{amsbsy}
\usepackage{setspace}
\usepackage[authoryear]{natbib}
\makeatletter

\providecommand{\tabularnewline}{\\}

\makeatother

\usepackage{babel}
\begin{document}

\title{Vaccines, Contagion, and Social Networks}

\author{Elizabeth L. Ogburn, Tyler J. VanderWeele%
\thanks{Elizabeth Ogburn (email: eogburn@jhsph.edu) is Assistant Professor,
Department of Biostatistics, Johns Hopkins University, Baltimore,
MD 21205; Tyler VanderWeele is Professor, Departments of Epidemiology
and Biostatistics, Harvard School of Public Health, Boston, MA 02115.
Dr. Ogburn\textquoteright{}s research was supported by grants U54
GM088558 and ES017678 from the National Institutes of Health. Dr.
VanderWeele's research was supported by grant ES017678 from the National
Institutes of Health.%
}}

\date{~}
\maketitle
\begin{abstract}
Consider the causal effect that one individual's treatment may have
on another individual's outcome when the outcome is contagious, with
specific application to the effect of vaccination on an infectious
disease outcome. The effect of one individual's vaccination on another's
outcome can be decomposed into two different causal effects, called
the ``infectiousness'' and ``contagion'' effects. We present identifying
assumptions and estimation or testing procedures for infectiousness
and contagion effects in two different settings: (1) using data sampled
from independent groups of observations, and (2) using data collected
from a single interdependent social network. The methods that we propose
for social network data require fitting generalized linear models
(GLMs). GLMs and other statistical models that require independence
across subjects have been used widely to estimate causal effects in
social network data, but, because the subjects in networks are presumably
not independent, the use of such models is generally invalid, resulting
in inference that is expected to be anticonservative. We introduce
a way to ensure that GLM residuals are uncorrelated across subjects
despite the fact that outcomes are non-independent. This simultaneously
demonstrates the possibility of using GLMs and related statistical
models for network data and highlights their limitations. 
\end{abstract}

\section{Introduction}

We are concerned here with the effect that one individual's treatment
may have on another individual's outcome, when the outcome is contagious.
In the infectious disease literature, this is often called an \textit{indirect
effect} of treatment \citep{halloran1991study}, while the effect
of an individual's treatment on his own outcome is a \textit{direct
effect}. Indirect effects of infectious disease interventions are
of significant importance for understanding infectious disease dynamics
and for designing public health interventions. For example, the goal
of many vaccination programs is to achieve herd immunity, whereby
a large enough subset of a population is vaccinated that even those
individuals who remain unvaccinated are protected against infection.
This is one type of indirect effect of a vaccination program; it has
been extensively studied in the infectious disease literature \citep{anderson1985vaccination,fine1993herd,john2000herd,o2003potential}.
Recently, interest has turned towards the identification and estimation
of average individual-level indirect effects \citep{halloran1991study,halloran1995causal,halloran2011causal,vanderweele2011bounding,vanderweele2011components,vanderweele2011effect},
such as the effect on a single member of a community of two different
vaccination programs implemented on the rest of the community \citep{halloran1995causal}. 

\citet{vanderweele2011components} demonstrated that the individual-level
indirect effect of vaccination in communities of size two can be decomposed
into to two different effects, called the ``infectiousness'' and
``contagion'' effects. These two effects represent distinct causal
pathways by which one person's vaccination may affect another's disease
status. The contagion effect is the indirect effect that vaccinating
one individual may have on another by preventing the vaccinated individual
from getting the disease and thereby from passing it on. The infectiousness
effect is the indirect effect that vaccination might have if, instead
of preventing the vaccinated individual from getting the disease,
it renders the disease less infectious, thereby reducing the probability
that the vaccinated infected individual transmits the disease, even
if infected.

\citet{vanderweele2011components} only considered estimation of the
infectiousness and contagion effects in a sample comprised of independent
households of size two with one member of each household assumed to
be homebound. The assumption that one individual is homebound and
the assumption of independent households are restrictive, the latter
because it requires that the households be sampled from distinct communities
and geographic areas. \citet{ogburnDAGs} considered the setting in
which households are independent but both individuals may be exposed
outside the household. Here, we relax the requirement of independent
households of size two and provide extensions to independent groups
of arbitrary size and to social networks. 

\textcolor{black}{Increasingly, data are available on the spread of
contagious outcomes through social networks. This setting is considerably
more complex than that considered in \citet{vanderweele2011components},
because the observed outcomes (e.g. disease status) are not independent
of one another. There is a growing literature on the possibility of
testing for the presence of different causal mechanisms using observational
data from social networks and a consensus that more rigorous methods
are needed. }

\textcolor{black}{An emerging body of work reports results from generalized
linear models (GLMs) and, for longitudinal data, generalized estimating
equations (GEEs) as estimates of peer effects, or the causal effect
that one individual's outcome may have on his or her social contacts'
outcomes \citep{ali2009estimating,cacioppo2009alone,christakis2007spread,christakis2008collective,christakis2013social,fowler2008estimating,lazer2010coevolution,rosenquist2010spread}.
This work has come under criticism that can largely be summarized
into two overarching themes. First, much of the criticism focuses
on the ability to control for confounding when estimating peer effects,
and specifically on the identifying assumptions that are required
in order to tell the difference between the well known problem of
homophily (the phenomenon by which individuals with similar traits
are more likely to form social ties with one another) and peer influence
\citep{cohen2008obesity,lyons2010spread,noel2011unfriending,shalizi2011homophily,vanderweele2011sensitivity}.
Homophily will not be an issue in many infectious disease settings,
as many such illnesses, for example the seasonal flu, are unlikely
to change the nature of social ties. Adequate control for confounding
is still crucial, but we assume throughout that all potential confounders
of the causal effects of interest are observed. This assumption should
be assessed in any application of these methods and it may not hold
in many real data settings; however, we do not focus on this assumption
in the remainder of this paper. }

\textcolor{black}{The second class of criticisms addresses the use
of statistical models for independent observations in this dependent
data setting. \citet{lyons2010spread} and \citet{vanderweele2012and}
demonstrated the importance of ensuring that models are coherent when
an observation can be both an outcome and a predictor (of social contacts'
outcomes); this is easily accomplished by using the observations at
one time point as predictors and the observations at a subsequent
time point as outcomes, a solution that was implemented in many of
applications of GLMs and GEEs to social network data referenced above.
More challenging is the fact that, when an analysis assumes independence
but observations are in fact positively correlated, as we would expect
them to be for contagious outcomes in a social network, the resulting
standard errors and statistical inference will generally be anticonservative.
}In some cases, the assumption of independent outcomes may hold under
the null hypothesis \citep{vanderweele2012and}, but it is unknown
whether tests that rely on this fact have any power to detect the
presence of the causal effects of interest \citep{shalizi2012statistics}. 

\textcolor{black}{Our contribution to methodology for social network
analysis is to adapt GLMs to ensure that the models can be correctly
specified, with uncorrelated residuals, even when the outcome is contagious.
We demonstrate the possibility of testing for the presence of contagion
and infectiousness effects using social network data and generalized
linear models (GLMs).} We discuss the paradigmatic example of the
effect of a vaccination on an infectious disease outcome, but effects
like contagion and infectiousness are of interest in other settings
as well. Our general approach to correctly specifying GLMs for a contagious
outcome using network data could potentially be applied to any estimand
for which GLMs are appropriate under independence. The\textcolor{black}{{}
tests that we propose have important limitations, most notably low
power to detect effects unless networks are large and/or sparse. However,
this work represents an important proof of concept in the ongoing
endeavor to develop methods for valid inference using data collected
from a single network. Furthermore, it clarifies the issues of model
misspecification and invalid standard errors raised by previous proposals
for using GMLs to assess peer effects using network data.}

\section{Social networks and contagion}

Formally, a social network is a collection of individuals and the
ties between them. The presence of a tie between two individuals indicates
that the individuals share some kind of a relationship; what types
of relationships are encoded by network ties depends on the context.
For example, we might define a network tie to include familial relatedness,
friendship, and shared place of work. Some types of relationships
are mutual, for example familial relatedness and shared place of work.
Others, like friendship, may go in only one direction: Tom may consider
Sue to be his friend, while Sue does not consider Tom to be her friend.
We will assume that all ties in our network are mutual or undirected,
but the principles of our method extend to directed ties. A node whose
characteristics we wish to explain is called an \textit{ego}; nodes
that share ties with the ego are its \textit{alters} or \textit{contacts}.
If an ego's outcome may be affected by his contacts' outcomes, then
we say that the outcome exhibits \textit{induction} or \textit{contagion}. 

Social networks are crucial to understanding many features of infectious
disease dynamics, and, increasingly, infectious disease researchers
draw on social network data to refine their understanding of transmission
patterns and treatment effects. For example, many mathematical models
of infectious disease now incorporate social network structure, whereas
they previously generally assumed uniform mixing among members of
a community \citep{eubank2004modelling,klovdahl1985social,klovdahl1994social,keeling2005networks},
and researchers collect data on sexual contact networks, since properties
of these networks can inform strategies for controlling sexually transmitted
diseases \citep{latora2006network,eames2002modeling,eames2004monogamous}. 

It is desirable for a number of reasons to study infectiousness and
contagion in the context of social networks rather than in independent
communities. First, social network data may be easier to collect or
to access than data on independent communities, as the latter setting
requires sampling from a large number of different locations or contexts
that are separated by time or space. Second, assessing whether traits
can be transmitted from one individual to another through network
ties is one of the central questions in the study of social networks;
assessing infectiousness and contagion contributes further insight
into this problem. Finally, social network data more realistically
capture the true interdependencies of the individuals whom we can
hope to treat with any public health intervention. Vaccine programs
do not in general target distant, independent pairs of individuals;
they target villages, cities, or communities in which individuals
are interconnected and their outcomes correlated. Therefore, assessing
the presence of vaccine effects in social network data may be more
informative for real-world applications. The methods we present here
represent a first step towards being able to estimate and perform
inference about such effects using social network data. 

Interventions to prevent infectious diseases generally operate in
two ways. Some reduce the susceptibility of treated individuals to
the disease, thereby preventing them from becoming infected. Examples
of such interventions are vaccines for tetanus, hepatitis A and B,
rabies, and measles \citep{keller2000passive}. These vaccines have
indirect effects that operate via contagion effects. Other interventions
may reduce the likelihood that an infected individual passes on his
infection to others. The malaria transmission-blocking vaccine is
designed to prevent mosquitos from acquiring, and thereby from transmitting,
malaria parasites upon biting infected individuals \citep{halloran1992modeling}.
This vaccine has no protective effect for the vaccinated individual,
but it renders vaccinated individuals less likely to transmit the
disease. Therefore any indirect effect of the malaria transmission-blocking
vaccine is due entirely to an infectiousness effect. Many interventions
have indirect effects that operate via both contagion and infectiousness
effects. 

Existing methods for assessing causal effects using network data are
limited. Some recent proposals give methods for assessing indirect
effects when treatment can be randomized \citep{airoldi2013estimation,aronow2012estimating,bowers2013reasoning,rosenbaum2007interference},
but these methods are of limited use in observational settings or
for teasing apart specific types of indirect effects like the infectiousness
and contagion effects. Much of the extant literature relies on GLMs
and GEEs, despite the fact that the key assumption of independent
outcomes across subjects is unlikely to hold in social network settings
\citep{lyons2010spread}. In this paper, we introduce a way to ensure
that GLM residuals are uncorrelated across subjects despite the fact
that outcomes are non-independent; this facilitates the use of GLMs
to assess infectiousness and contagion effects in social network contexts.
We demonstrate through simulations that our methods do have some power
to detect the presence of contagion and infectiousness effects; however,
in order to ensure that residuals are uncorrelated, we make several
adaptations to naive GLMs; unfortunately these can result in low power.
The applications that we discuss in this paper do not require the
use of GEEs to account for within-subject dependence over time, but
the general principles that we use to adapt GLMs to the network setting
apply to GEEs as well.

\section{Infectiousness and contagion in independent groups of size $2$}

\subsection{Notation and assumptions }

Consider $K$ households comprised of two individuals each and separated
by space or time such that an infectious disease cannot be transmitted
between individuals in different households. Borrowing terminology
from the social network literature, we will refer to one individual
as the alter, denoted $a$, and the other as the ego, denoted $e$.
For now we assume that in each household the ego is unvaccinated,
and that all vaccination occurs before the start of follow-up. Contagion
and infectiousness effects are analogous to causal mediation effects
of the alter's vaccination on the ego's outcome, mediated by the alter's
disease status \citep{vanderweele2011components}. We formally define
these effects in the next section after first introducing key notation
and identifying assumptions. 

For individual $i$ in household $k$, $i=a,e$, let $Y_{i_{k}}^{t}$
be the outcome at time $t$ and $C_{i_{k}}$ be a vector of covariates.
Let $V_{a_{k}}$ be an indicator of vaccination for the alter in household
$k$. Below we omit the subscript $k$ when context allows. Define
$Y_{i_{k}}^{t}(v)$ to be the counterfactual outcome we would have
observed for individual $i$ in household $k$ at time $t$, if, possibly
contrary to fact, the alter had received treatment $v$. Let $M_{k}$
be a variable that lies on a causal pathway from $V_{a_{k}}$ to $Y_{e_{k}}^{t}$.
Let $Y_{e_{k}}^{t}(v,m)$ be the counterfactual outcome for the ego
at time $t$ that we would have observed if $V_{a_{k}}$ had been
set to $v$ and $M_{k}$ to $m$. Throughout we make the consistency
assumptions that $M_{k}(v)=M_{k}$ when $V_{a_{k}}=v$, that $Y_{e_{k}}(v,m)=Y_{e_{k}}$
when $V_{k}=v$ and $M_{k}=m$, and that $Y_{e_{k}}(v,M_{k}(v))=Y_{e_{k}}(v)$.
Let $Y_{e_{k}}^{t}\left(v,M_{k}(v')\right)$ be the counterfactual
disease status for the ego in household $k$ that we would have observed
at time $t$ if $V_{a_{k}}$ had been set to $v$ and $M_{k}$ to
its counterfactual value under $V_{a_{k}}=v'$. To ensure that this
counterfactual is well-defined, we assume that it is hypothetically
possible to intervene on the mediator without intervening on $V_{a_{k}}$.
Let $\boldsymbol{C}_{k}=(C_{a_{k}},C_{e_{k}})$. In order to identify
functionals of nested counterfactuals like $Y_{e}^{t}\left(v,M(v')\right)$
we require the following four assumptions \citep{pearl2001direct}:
\begin{equation}
\left.Y_{e}^{t}\left(v,m\right)\perp V_{a}\right|\boldsymbol{C},\label{eq:med1}
\end{equation}
\begin{equation}
\left.Y_{e}^{t}\left(v,m\right)\perp M\right|V_{a},\boldsymbol{C},\label{eq:med2}
\end{equation}
\begin{equation}
\left.M\left(v\right)\perp V_{a}\right|\boldsymbol{C},\label{med3}
\end{equation}
and 
\begin{equation}
\left.Y_{e}^{t}\left(v,m\right)\perp M\left(v'\right)\right|\boldsymbol{C}\label{med4}
\end{equation}
where $A\perp B\mid C$ denotes that $A$ is independent of $B$ conditional
on $C$. Assumptions (\ref{eq:med1}), (\ref{eq:med2}), and (\ref{med3})
correspond to the absence of unmeasured confounders for the effects
of the exposure on the outcome ($V_{a}$ on $Y_{e}^{t}$ ), of the
mediator on the outcome ($M$ on $Y_{e}^{t}$), and of the exposure
on the mediator ($V_{a}$ on $M$), respectively. Assumption (\ref{med4})
requires that no confounder of the effect of $M$ on $Y_{e}^{t}$
is affected by $V_{a}$$ $. Discussion of these assumptions in the
context of mediation analysis can be found in \citet{pearl2001direct}.
Discussion and extension of these assumptions to settings with interference
or spillover effects can be found in \citet{ogburnDAGs}, including
discussion of how to determine which covariates must be included in
$\boldsymbol{C}$.

\subsection{Previous methodology for decomposing the indirect effect into infectiousness
and contagion effects}

\subsubsection{Identification}

\citet{vanderweele2011components} described the decomposition of
indirect effects into contagion and infectiousness effects in communities
of size two. They assumed that the outcome can only occur once for
each individual during the follow-up period. This is a reasonable
assumption for many infectious disease outcomes, for example for the
common flu with a follow-up period consisting of a single flu season.
They further assumed that in each pair the ego cannot be exposed to
the disease except by the alter, as might be the case if the ego were
homebound. Let $t_{f}$ be the time of the end of follow-up and $Y_{ke}^{t_{f}}$
be an indicator of whether the ego in household $k$ has had the disease
by the end of follow-up. \citet{vanderweele2011components} defined
the population average indirect effect of vaccination on the ego as
$E\left[Y_{e}^{t_{f}}(1)\right]-E\left[Y_{e}^{t_{f}}(0)\right]$,
or the expected difference in the counterfactual disease status of
the ego at end of follow-up when the alter is vaccinated compared
to when the alter is not vaccinated. When the ego is homebound, the
indicator $Y_{ke}^{t_{f}}$ of whether the ego in household $k$ has
had the disease by the end of follow-up is, equivalently, an indicator
of whether the ego was infected by the alter in household $k$. In
order to generalize the discussion of vaccine effects to settings
in which the ego can be infected from outside the home, the outcome
$Y_{ke}^{t_{f}}$ should be defined more precisely as the indicator
of whether the ego was sick after the alter. Specifically, let $Y_{ke}^{t_{f}}=I(\mbox{alter was sick at time }T<t_{f}\mbox{ and ego was sick at time \ensuremath{S,\, T<S\leq t_{f}}})$. 

The contagion effect is the protective effect that treating one individual
has on another's disease status by preventing the treated individual
from getting the disease and thereby from transmitting it. Let $T_{k}$
be the time of the first case of the disease in household $k$. This
is akin to the effect of one individual's treatment on another's disease
status as mediated by the first individual's disease status. For the
purposes of the analysis below, we define a disease case to begin
when an individual becomes infectious. If infectiousness does not
coincide with the appearance of disease symptoms then we may not observe
the timing of disease cases directly, but we could infer the time
based on when symptoms appear and on known disease dynamics. For example,
an individual with the flu will generally be infectious one day before
he is symptomatic \citep{earn2002ecology}. Therefore, if flu is the
disease under study we would classify an individual as having the
disease beginning one day before he reported having flu symptoms.
We assume throughout that there are no asymptomatic carriers of the
disease. If neither individual in household $k$ is ever sick then
we define $T_{k}$ to be the end of follow-up. Now $Y_{a_{k}}^{T_{k}}$
is an indicator of whether the alter is sick at time $T_{k}$, i.e.
an indicator of whether the alter is the first individual in the group
to get sick; if neither individual gets sick then it will be 0. Let
$T_{k}(v)$ be the time at which the first infection in household
$k$ would have occurred if the alter had, possibly contrary to fact,
had vaccine status $v$. Let $Y_{a_{k}}^{T_{k}(v)}(v)$ be the counterfactual
disease status of the alter at time $T_{k}(v)$ had he had vaccine
status $v$. Let $Y_{e_{k}}^{t_{f}}=I(\mbox{individual \ensuremath{e_{k}}became infectious after time \ensuremath{T_{k}}}\mbox{and on or before time }t_{f})$.
The contagion effect is given by a contrast in counterfactuals of
the form $Y_{e}^{t_{f}}\left(v,Y_{a}^{T(v')}(v')\right)$ where, unlike
in the mediation framework we described in Section 3.1, the variable
$Y_{a}^{T(v')}$ that plays the role of mediator may be a different
random variable in the two terms in the contrast. Specifically, the
population average contagion effect is $E\left[Y_{e}^{t_{f}}\left(0,Y_{a}^{T(1)}(1)\right)\right]-E\left[Y_{e}^{t_{f}}(0,Y_{a}^{T(0)}(0))\right]$,
and $Y_{a}^{T(0)}$ and $Y_{a}^{T(1)}$ will be different random variables
whenever $T(0)\neq T(1)$. This contrast is the difference in expected
counterfactual outcomes for the ego when the vaccine status of the
alter is held constant at 0 but his infection status is set to that
under vaccination in the first term and to that under no vaccination
in the second term of the contrast. It captures the effect that vaccination
might have had on the disease status of the ego by preventing the
alter from contracting the disease. The nested counterfactuals are
well-defined because we can imagine intervening on $Y_{ka}^{T_{k}}$
without intervening on $V_{ka}$, for example by administering immune
boosters to prevent the alter from being infected or by exposing the
alter to a high dose of flu virus in a laboratory setting to cause
infection. 

The population average infectiousness effect is $E\left[Y_{e}^{t_{f}}\left(1,Y_{a}^{T(1)}(1)\right)\right]-E\left[Y_{e}^{t_{f}}(0,Y_{a}^{T(1)}(1))\right]$.
This is akin to the effect of one individual's treatment on another's
disease status, not mediated through the first individual's disease
status. This effect operates if treatment renders cases of disease
among treated individuals less likely to be transmitted. Suppose that
the alter in group $k$ would get the flu first if vaccinated. That
is, $Y_{a_{k}}^{T_{k}(1)}(1)=1$. Then the infectiousness effect is
the difference in counterfactual outcomes for the ego comparing the
scenario in which the alter is vaccinated and infected first with
the scenario in which to the alter is unvaccinated and infected first.
If the alter in group $k$ would not get the flu first under vaccination,
then the infectiousness effect for group $k$ is null.

By the consistency assumption we made in Section 3.1 above, $E\left[Y_{e}^{t_{f}}\left(1,Y_{a}^{T(1)}(1)\right)\right]=E\left[Y_{e}^{t_{f}}\left(1\right)\right]$
and $E\left[Y_{e}^{t_{f}}\left(0,Y_{a}^{T(0)}(0)\right)\right]=E\left[Y_{e}^{t_{f}}\left(0\right)\right]$.
The indirect effect of the vaccination of the alter on the ego decomposes
into the sum of the contagion and infectiousness effects as follows:
\begin{eqnarray*}
 &  & E\left[Y_{e}^{t_{f}}(1)\right]-E\left[Y_{e}^{t_{f}}(0)\right]\\
 & = & E\left[Y_{e}^{t_{f}}\left(1,Y_{a}^{T(1)}(1)\right)\right]-E\left[Y_{e}^{t_{f}}\left(0,Y_{a}^{T(0)}(0)\right)\right]\\
 & = & E\left[Y_{e}^{t_{f}}\left(1,Y_{a}^{T(1)}(1)\right)\right]-E\left[Y_{e}^{t_{f}}(0,Y_{a}^{T(1)}(1))\right]+E\left[Y_{e}^{t_{f}}\left(0,Y_{a}^{T(1)}(1)\right)\right]-E\left[Y_{e}^{t_{f}}(0,Y_{a}^{T(0)}(0))\right]
\end{eqnarray*}

The assumptions made by \citet{vanderweele2011components} allow for
the identification of the infectiousness and contagion effects even
if disease status is only observed at the end of follow-up. Because
the ego cannot be infected except by the alter, $Y_{a_{k}}^{T_{k}}=0$
if and only if neither individual is observed to get sick and $Y_{a_{k}}^{T_{k}}=1$
if and only if $Y_{a_{k}}^{t_{f}}=1$. Therefore $Y_{a_{k}}^{t_{f}}$
can be substituted for $Y_{a_{k}}^{T_{k}}$ in the expressions above. 

\citet{ogburnDAGs} gave identifying assumptions for the infectiousness
and contagion effects in groups of size two when the time of infections
is observed. They did not assume that only one member of each pair
is exposed from outside of the group; instead they assumed that the
probability of the ego contracting the disease within a fixed follow-up
interval if exposed at time $t$ is constant in $t$. This ensures
that the time of the first infection $T$ is not a confounder of the
mediator-outcome relationship, which would constitute a violation
of assumption (\ref{med4}) because $T$ is affected by $V_{a}$.
Suppose that the two members of each pair are distinguishable from
one another, for example parent-child pairs. We select one of the
two to be the alter (e.g the parent) and the other is the ego (the
child). Alternatively, if the individuals are exchangeable, that is,
if we have no reason to think that the indirect effect and its components
will be different for one than for the other, then we can randomly
choose which subject is the alter and which is the ego. \citet{ogburnDAGs}
defined the indicator $Y_{e}^{T+s}$ of whether the ego is sick after
time $T$ and by time $T+s$ to be the outcome, where $s$ is a constant
that allows $T$ to determine a new end of follow-up. This ensures
that $T$ does not confound the mediator-outcome relationship. The
constant $s$ should be chosen to be the sum of the infectious period
($f$) and the incubation period ($b$) of the disease under study.
The infectious period is the length of time during which an infected
individual is infectious, and the incubation period is the length
of time between being infected and becoming infectious. If the alter
becomes infectious at time $T$, then he can infect the ego until
time $T+f$. If infected at time $T+f$, the ego will become infectious
at time $T_{k}+f+b=T_{k}+s$. Therefore if the alter infects the ego,
the ego must be infectious by time $T_{k}+s$. We assume throughout
that the time to efficacy of vaccine is immediate and that the infectious
and incubation periods are constant across individuals. 

Let $Y_{e_{k}}^{T_{k}(v')+s}\left(v,Y_{a_{k}}^{T_{k}(v')}(v')\right)$
be the counterfactual outcome we would have observed for the ego in
group $k$ at time $T_{k}(v')+s$ if the alter's vaccine status were
set to $v$ and the alter's disease status at time $T_{k}(v')$ were
set to its counterfactual under vaccine status $v'$. The average
contagion effect in this setting is given by $E\left[Y_{e}^{T(1)+s}\left(0,Y_{a}^{T(1)}(1)\right)\right]$
$-$ $E\left[Y_{e}^{T(0)+s}\left(0,Y_{a}^{T(0)}(0)\right)\right]$
and the average infectiousness effect by $E\left[Y_{e}^{T(1)+s}\left(1,Y_{a}^{T(1)}(1)\right)\right]$
$-$ $E\left[Y_{e}^{T(1)+s}\left(0,Y_{a}^{T(1)}(1)\right)\right]$.
The sum of these two effects is the average indirect effect $E\left[Y_{e}^{T(1)+s}(1)\right]-E\left[Y_{e}^{T(0)+s}(0)\right]$.
Although the disease status of the ego is measured $s$ days after
the first infection instead of at the end of follow-up, this indirect
effect still captures any effect that the alter's vaccination status
can have on the ego's disease status, because after time $T+s$ any
change in the disease status of the ego cannot be caused by $V_{a}$. 

So far we have described all effects on the difference scale, but
everything we have written applies equally to effects on the ratio
and odds ratio scales. On the ratio and odds ratio scales the indirect
effect of vaccination decomposes into a product of the contagion and
infectiousness effects. On the ratio scale, the average indirect effect
of $V_{a}$ on the disease status of the ego is $E\left[Y_{e}^{T(1)+s}(1)\right]/E\left[Y_{e}^{T(0)+s}(0)\right]$,
which is a product of the average infectiousness effect, $E\left[Y_{e}^{T(1)+s}\left(1,Y_{a}^{T(1)}(1)\right)\right]/E\left[Y_{e}^{T(1)+s}(0,Y_{a}^{T(1)}(1))\right]$,
and the average contagion effect, $E\left[Y_{e}^{T(1)+s}\left(0,Y_{a}^{T(1)}(1)\right)\right]/E\left[Y_{e}^{T(0)+s}(0,Y_{a}^{T(0)}(0))\right]$.
On the odds ratio scale for a binary outcome the decomposition is
\begin{eqnarray*}
 &  & \frac{E\left[Y_{e}^{T(1)+s}(1)\right]\left(1-E\left[Y_{e}^{T(0)+s}(0)\right]\right)}{E\left[Y_{e}^{T(0)+s}(0)\right]\left(1-E\left[Y_{e}^{T(1)+s}(1)\right]\right)}\\
 & = & \frac{E\left[Y_{e}^{T(1)+s}\left(1,Y_{a}^{T(1)}(1)\right)\right]\left(1-E\left[Y_{e}^{T(1)+s}(0,Y_{a}^{T(1)}(1))\right]\right)}{E\left[Y_{e}^{T(1)+s}(0,Y_{a}^{T(1)}(1))\right]\left(1-E\left[Y_{e}^{T(1)+s}\left(1,Y_{a}^{T(1)}(1)\right)\right]\right)}\\
 &  & \times\frac{E\left[Y_{e}^{T(1)+s}\left(0,Y_{a}^{T(1)}(1)\right)\right]\left(1-E\left[Y_{e}^{T(0)+s}(0,Y_{a}^{T(0)}(0))\right]\right)}{E\left[Y_{e}^{T(0)+s}(0,Y_{a}^{T(0)}(0))\right]\left(1-E\left[Y_{e}^{T(1)+s}\left(0,Y_{a}^{T(1)}(1)\right)\right]\right)}
\end{eqnarray*}
where the first line is the indirect effect, the second line is the
infectiousness effect, and the third line is the contagion effect$ $.

\subsubsection{Estimation}

The contagion and infectiousness effects are analogous to the natural
indirect and direct effects, respectively, of the effect of $V_{a}$
on $Y_{e}^{T+s}$ with $Y_{a}^{T}$ as the mediator. Natural indirect
and direct effects have been written about extensively in the causal
inference and mediation literature (see e.g. \citealp{pearl2001direct,robins1992identifiability,RobinsRichardsonBook})
and it is well-known how to estimate them in a variety of settings
\citep{imai2010general,valeri2012}. This setting differs from those
considered by other authors because the outcome $Y_{e}^{T+s}$ is,
by definition, equal to $0$ whenever $Y_{a}^{T}$ is equal to $0$;
therefore one must be careful to ensure that any model specified for
for $E\left[Y_{e}^{T+s}\mid V_{a},Y_{a}^{T},\boldsymbol{C}\right]$
is consistent with this restriction. \citet{vanderweele2011components}
describe how to estimate the contagion and infectiousness effects
on the ratio scale in households of size two when one individual is
homebound, but the procedure they present overlooks this restriction
and therefore the models they suggest may fail to converge. 

We describe a procedure for estimating the contagion and infectiousness
effects that is appropriate for the setting considered in \citet{vanderweele2011components}
and for the setting in which neither individual is assumed to be homebound.
We describe estimation of the effects on the difference and ratio
scales. Estimation of effects on the odds ratio scale is also possible.
Suppose that assumptions (\ref{eq:med1}) through (\ref{med4}) hold
for the effect of $V_{a}$ on $Y_{e}^{T+s}$ with $Y_{a}^{T}$ as
the mediator and covariates $\boldsymbol{C}$, and that the following
two models are correctly specified: 
\begin{eqnarray}
log\left\{ E\left[Y_{e}^{T+s}\mid V_{a},Y_{a}^{T}=1,\boldsymbol{C}\right]\right\}  & = & \gamma_{0}+\gamma_{1}V_{a}+\gamma_{2}'\boldsymbol{C}\label{model1}\\
logit\left\{ E\left[Y_{a}^{T}\mid V_{a},\boldsymbol{C}\right]\right\}  & = & \eta_{0}+\eta_{1}V_{a}+\eta_{2}'\boldsymbol{C}.\label{model2}
\end{eqnarray}
If the outcome is rare then (\ref{model1}) can be replaced with a
logistic model. The contagion effect conditional on covariates $\boldsymbol{C}=\boldsymbol{c}$
on the difference scale is given by 
\begin{eqnarray*}
 &  & E\left[Y_{e}^{T(1)+s}(0,Y_{a}^{T(1)}(1))\mid\boldsymbol{c}\right]-E\left[Y_{e}^{T(0)+s}(0,Y_{a}^{T(0)}(0))\mid\boldsymbol{c}\right]\\
 &  & =0+E\left[Y_{e}^{T+s}\mid V_{a}=0,Y_{a}^{T}=1,\boldsymbol{c}\right]\left\{ E\left[Y_{a}^{T}\mid V_{a}=1,\boldsymbol{c}\right]-E\left[Y_{a}^{T}\mid V_{a}=0,\boldsymbol{c}\right]\right\} \\
 &  & =e^{\gamma_{0}+\gamma_{2}'\boldsymbol{c}}\left\{ \frac{e^{\eta_{0}+\eta_{1}V_{a}+\eta_{2}'\boldsymbol{c}}}{1+e^{\eta_{0}+\eta_{1}V_{a}+\eta_{2}'\boldsymbol{c}}}-\frac{e^{\eta_{0}+\eta_{2}'\boldsymbol{c}}}{1+e^{\eta_{0}+\eta_{2}'\boldsymbol{c}}}\right\} .
\end{eqnarray*}
and the infectiousness effect conditional on covariates $\boldsymbol{C}=\boldsymbol{c}$
is given by 
\begin{eqnarray*}
 &  & E\left[Y_{e}^{T(1)+s}(1,Y_{a}^{T(1)}(1))\mid\boldsymbol{c}\right]-E\left[Y_{e}^{T(1)+s}(0,Y_{a}^{T(1)}(1))\mid\boldsymbol{c}\right]\\
 &  & =0+E\left[Y_{a}^{T}\mid V_{a}=1,\boldsymbol{c}\right]\left\{ E\left[Y_{e}^{T+s}\mid V_{a}=1,Y_{a}^{T}=1,\boldsymbol{c}\right]-E\left[Y_{e}^{T+s}\mid V_{a}=0,Y_{a}^{T}=1,\boldsymbol{c}\right]\right\} \\
 &  & =\frac{e^{\eta_{0}+\eta_{1}V_{a}+\eta_{2}'\boldsymbol{c}}}{1+e^{\eta_{0}+\eta_{1}V_{a}+\eta_{2}'\boldsymbol{c}}}\left\{ e^{\gamma_{0}+\gamma_{1}V_{a}+\gamma_{2}'\boldsymbol{c}}-e^{\gamma_{0}+\gamma_{2}'\boldsymbol{c}}\right\} .
\end{eqnarray*}

The contagion and infectiousness effects can be estimated by fitting
models (\ref{model1}) and (\ref{model2}) and plugging the parameter
estimates into the expressions above. The standard errors for these
estimates can be bootstrapped or derived using the delta method (similar
to those derived in \citealp{valeri2012} for the natural direct and
indirect effects). Alternatively, a Monte Carlo based approach similar
to \citet{imai2010general} can be used for estimation of the effects
and their standard errors. Software packages like SAS and SPSS mediation
macros \citep{valeri2012} or the R mediation package \citep{imai2010general}
cannot be used in this setting because instead of (\ref{model1}),
which models the conditional expectation of the $Y_{e}^{T+s}$ only
in the $Y_{a}^{T}=1$ stratum, these packages require fitting a model
for $ $ $E\left[Y_{e}^{T+s}\mid V_{a},Y_{a}^{T},\boldsymbol{C}\right]$.

If the ego can also be vaccinated then $V_{e}$ must be included in
$\boldsymbol{C}$. If $V_{a}$ interacts with $V_{e}$ or with any
other covariates, these interactions can be incorporated into the
models and pose no difficulty for estimation. To test whether there
is a contagion effect, we can simply test whether $\eta_{1}=0$. To
test whether there is an infectiousness effect we can simply test
whether $\gamma_{1}=0$. 

Using the parameters of models (\ref{model1}) and (\ref{model2})
we can also estimate the contagion and infectiousness effects on the
ratio scale. The contagion effect conditional on $\boldsymbol{C}=\boldsymbol{c}$
is given by 
\begin{eqnarray}
\frac{E\left[Y_{e}^{T(1)+s}(0,Y_{a}^{T(1)}(1))\mid\boldsymbol{c}\right]}{E\left[Y_{e}^{T(0)+s}(0,Y_{a}^{T(0)}(0))\mid\boldsymbol{c}\right]} & = & \frac{0+E\left[Y_{e}^{T+s}\mid V_{a}=0,Y_{a}^{T}=1,\boldsymbol{c}\right]E\left[Y_{a}^{T}\mid V_{a}=1,\boldsymbol{c}\right]}{0+E\left[Y_{e}^{T+s}\mid V_{a}=0,Y_{a}^{T}=1,\boldsymbol{c}\right]E\left[Y_{a}^{T}\mid V_{a}=0,\boldsymbol{c}\right]}\nonumber \\
 & = & \frac{E\left[Y_{a}^{T}\mid V_{a}=1,\boldsymbol{c}\right]}{E\left[Y_{a}^{T}\mid V_{a}=0,\boldsymbol{c}\right]}\nonumber \\
 & = & \frac{e{}^{\eta_{1}}+e^{\eta_{0}+\eta_{1}+\eta_{2}'\boldsymbol{c}}}{1+e^{\eta_{0}+\eta_{1}+\eta_{2}'\boldsymbol{c}}}\label{con params}
\end{eqnarray}
and the infectiousness effect is given by 
\begin{eqnarray}
\frac{E\left[Y_{e}^{T(1)+s}(1,Y_{a}^{T(1)}(1))\mid\boldsymbol{c}\right]}{E\left[Y_{e}^{T(1)+s}(0,Y_{a}^{T(1)}(1))\mid\boldsymbol{c}\right]} & = & \frac{0+E\left[Y_{e}^{T+s}\mid V_{a}=1,Y_{a}^{T}=1,\boldsymbol{c}\right]E\left[Y_{a}^{T}\mid V_{a}=1,\boldsymbol{c}\right]}{0+E\left[Y_{e}^{T+s}\mid V_{a}=0,Y_{a}^{T}=1,\boldsymbol{c}\right]E\left[Y_{a}^{T}\mid V_{a}=1,\boldsymbol{c}\right]}\nonumber \\
 & = & \frac{E\left[Y_{e}^{T+s}\mid V_{a}=1,Y_{a}^{T}=1,\boldsymbol{c}\right]}{E\left[Y_{e}^{T+s}\mid V_{a}=0,Y_{a}^{T}=1,\boldsymbol{c}\right]}\label{inf params}\\
 & = & e^{\gamma_{1}}.\nonumber 
\end{eqnarray}
Under the restriction that $Y_{e}^{T+s}=0$ whenever $Y_{a}^{T}=0$,
the contagion effect on the ratio scale is simply a measure of the
effect of the alter's vaccination on the alter's outcome. It is mathematically
undefined if $E\left[Y_{e}^{T+s}\mid V_{a}=0,Y_{a}^{T}=1,\boldsymbol{c}\right]=0$,
that is, if the ego's outcome has no effect on the alter's outcome,
but it is natural to define it to be equal to the null value of $1$
in this case. The infectiousness effect on the ratio scale is simply
a measure of the effect of the alter's vaccination on the ego's outcome
among pairs in which the alter is sick first, that is, in the $Y_{a}^{T}=1$
stratum.

\section{Infectiousness and contagion in groups of more than two}

Although allowing both individuals in a household to be infected from
outside the household generalizes the results of \citet{vanderweele2011components},
it still requires the strong assumption, inherent in the identifying
assumptions described in Section 3, that the alter and ego do not
share any potentially infectious contacts. If both of the individuals
in a given household could be infected from outside the household
by the same mutual friend, then that friend's disease status would
be a confounder of the mediator-outcome relationship; if unobserved,
it would constitute a violation of assumption (\ref{eq:med2}). We
can relax the assumption of no mutual contacts outside of the household
by collecting data on any such contacts and controlling for them as
covariates in our estimating procedure. 

In this section, we consider identification and estimation of the
contagion and infectiousness effects when independent groups of individuals
are sampled. We assume that each group includes a pair of individuals
who furnish the exposure, mediator, and outcome variables, plus all
mutual and potentially infectious contacts of the pair. Several types
of sampling procedures could give rise to this data structure. For
example, one possibility would be to sample workplaces and randomly
select two individuals to play the role of the alter and ego; another
would be to sample household pairs first, ascertain the identities
of potential mutual contacts outside of the home, and include all
such contacts in the data collection moving forward. The sampling
procedure does not affect the identification or estimation results
described below. 

Let $k$ index the $k^{th}$ group, $k=1,...,K$. Let $Y_{i_{k}}^{t}$
be an indicator of whether individual $i$ in group $k$ has had the
disease by day $t$. As in Section 3, we define a case of the disease
to begin when the individual becomes infectious and let $s=f+b$ be
the sum of the infectious and incubation periods for the disease.
We assume that vaccination occurs before the start of follow-up. Given
a non-rare outcome like the flu and time measured in discrete intervals
like days, it is likely that we would observe multiple individuals
to get sick on the same day. We therefore do not make the assumption,
made in Section 3, that no two individuals can be observed to get
sick at the same time. For group $k$, let $e_{k}$ index the ego,
whose flu status we wish to study, and let $a_{k}$ index the alter,
whose vaccination status may or may not have an effect on the ego's
disease status. We index the other individuals in group $k$ by $1,2,...,n_{k}$.
Let $T_{k}$ be the time of the first infection in the $k^{th}$ alter-ego
pair. As in Section 3, the ego furnishes the outcome, $Y_{e_{k}}^{T_{k}+s}$.
The alter furnishes the treatment, vaccine status $V_{a_{k}}$, and
the mediator, indicator of first infection $Y_{a_{k}}^{T_{k}}$. When
context allows, we omit the subscript $k$. The definition of the
mediator needs to be modified slightly to reflect the fact that the
alter and the ego could get sick at the same time: let $Y_{a}^{T}$
be an indicator of whether the alter was sick and the ego healthy
at time $T$.\textbf{ }Let $Y_{e}^{T+s}$ be an indicator of whether
the ego got sick between time $T+b$, which is the first time at which
the alter could have infected the ego, and time $T+s$, which is the
last time at which the alter could have infected the ego. This definition
preserves the interpretation of $Y_{a}^{T}$ as an indicator that
the alter was sick before the ego; if the ego and the alter simultaneously
fell ill on day $T$ then $Y_{a}^{T}$ will be $0$, which is desirable
because the ego cannot have caught the disease from the alter if they
both fell ill on the same day. It also preserves the restriction,
discussed in Section 3, that $Y_{e}^{T+s}$ is equal to $0$ whenever
$Y_{a}^{T}$ is. 

$Y_{e}^{T(v')+s}\left(v,Y_{a}^{T(v')}(v')\right)$ is the counterfactual
flu status of the ego at time $T(v')+s$ had the alter's vaccine status
been set to $v$ and his flu status at time $T(v')$ set to its counterfactual
value under vaccine status $v'$, where $T(v')$ is the time at which
the first infection in the alter-ego pair would have occurred if $V_{a}$
had been set to $v'$. The effects of interest are the average contagion
effect 
\begin{equation}
Con=\frac{E\left[Y_{e}^{T(1)+s}\left(0,Y_{a}^{T(1)}(1)\right)\right]}{E\left[Y_{e}^{T(0)+s}\left(0,Y_{a}^{T(0)}(0)\right)\right]}\label{Con}
\end{equation}
and the average infectiousness effect 
\begin{equation}
Inf=\frac{E\left[Y_{e}^{T(1)+s}\left(1,Y_{a}^{T(1)}(1)\right)\right]}{E\left[Y_{e}^{T(1)+s}\left(0,Y_{a}^{T(1)}(1)\right)\right]},\label{Inf}
\end{equation}
where the expectations are taken over all ego-alter pairs.

In order to identify the effects defined in (\ref{Con}) and (\ref{Inf}),
we must measure and control for all confounders of the relationships
between $Y_{e}^{T+s}$ and $Y_{a}^{T}$, and in particular the potential
mutual infectious contacts of the alter and ego. To motivate our procedure
for controlling for these confounding contacts, consider the simple
case of a group of size three, comprised of a child (ego), a parent
(alter), and a grandparent. In the event that the grandparent contracted
the flu first and transmitted it to both the child and the parent,
the grandparent's flu status would clearly be a confounder of the
mediator-outcome relationship. But the grandparent's entire disease
trajectory is not a potential confounder; in particular anything that
happens to the grandparent after time $T$, that is after the first
infection in the parent-child pair, occurs after the mediator and
cannot possibly confound the mediator-outcome relationship. In this
simple, three-person group, it suffices to control for an indicator
of whether the grandparent has been sick by time $T-b$, where $T$
is the time of the first infection between the parent and child, and
$T-b$ is the latest time at which the grandparent could have been
the cause of an infection at time $T$. 

In practice, we will likely have to sample groups of size greater
than three in order to control for confounding by potential mutually
infectious contacts. It is generally sufficient to control for a summary
measure of the infections occurring before $T-b$ in each group. If
each infectious contact of an individual has an independent probability
of transmitting the disease to the individual, then the sum $\sum_{i=1}^{n_{k}}Y_{ki}^{T-b}$
of indicators of whether each mutual contact has been sick by time
$T-b$ suffices to control for confounding by potential mutual infectious
contacts. Under a different transmission model, the proportion $\sum_{i=1}^{n_{k}}Y_{ki}^{T-b}\left/n_{k}\right.$
of contacts who were sick by time $T-b$ could be the operative summary
measure. If some of the mutual contacts may have been vaccinated,
then separate summary measures (sum or proportion sick by time $T-b$)
should be included for vaccinated and for unvaccinated contacts. In
what follows we will assume that the sum is an adequate summary measure.

\subsection{Alternative sampling schemes}

Alter-centric sampling can also be used to collect data on variables
that suffice to identify the contagion and infectiousness effects.
Instead of sampling an alter-ego pair and all of their mutual contacts,
we can sample an individual to serve as the alter and all of his potentially
infectious contacts. The ego is randomly selected from among the alter's
contacts. Conditional on the number of the alter's contacts who have
been infectious by day $T-b$, $Y_{a}^{T}$ is independent of the
number of mutual contacts who were sick by time $T-b$. The number
of mutual contacts is no longer a confounder of the relationship between
$Y_{a}^{T}$ and $Y_{e}^{T+s}$ and there is no need to ascertain
the identity or disease status of the mutual contacts. However, the
number of potentially infectious contacts of a single person can be
vast, and it may be easier to identify mutual contacts of a pair of
individuals than all contacts of any one individual.

\section{Infectiousness and contagion in social networks}

So far, we have assumed that our observations, comprised of groups
of individuals, were independent of one another. This assumption will,
in general, be violated when the alter-ego pairs are sampled from
a single community or\textbf{ }social network. We introduce some new
notation for this context after briefly describing the example that
will serve as the basis for our exposition and later for our simulations
and data analysis. Consider tracking the seasonal flu in the student
population of a college at which all students live in dorms on campus.
Each student is a node in the network. We define a tie to exist between
two nodes if the individuals regularly interact with one another in
a way that could facilitate transmission of the flu. For example,
if two individuals are roommates, eat together in the dining hall,
or are close friends, then their nodes share a tie. We observe each
individual's flu status every day over the course of the flu season,
which lasts for 100 days. 

The contagion and infectiousness effects $Con$ and $Inf$, defined
in Section 4, are not estimable from social network data using the
methods that we propose below. Instead we can define new contagion
and infectiousness effects such that hypothesis tests based on the
new effects are valid and consistent tests of the hypotheses that
$Con$ and $Inf$ are null. We give assumptions under which the new
estimands are estimable from network data using GLMs and we demonstrate
that tests of the hypotheses for the new estimands are valid and consistent
for $Con$ and $Inf$.

\subsection{Assumptions}

Along with assumptions (\ref{eq:med1}) - (\ref{med4}), we make several
additional assumptions that facilitate inference using social network
data. Define $\mathcal{A}_{i}=\left\{ j:\mbox{ }i\mbox{ and }j\mbox{ share a tie}\right\} $
to be the collection of indices for individual $i$'s contacts. We
assume that 
\begin{equation}
Y_{i}^{t}\perp Y_{j}^{r}\mid\left\{ \sum_{m\in\mathcal{A}_{i}:V_{m}=v}Y_{m}^{t-b},\, v=0,1\right\} ,\,\mbox{for all }j\notin\mathcal{A}_{i}\mbox{ and }r\leq t.\label{info barrier}
\end{equation}
The set in the conditioning event includes the number of vaccinated
contacts of individual $i$ who were sick on or before day $t-b$
and the number of unvaccinated contacts of individual $i$ who were
sick on or before day $t-b$. This assumption says that the outcome
of individual $i$ at time $t$ is independent of all past outcomes
for non-contacts of $i$, conditional on a summary measure of the
flu history of the contacts of $i$. In other words, contacts act
as a causal barrier between two nodes who do not themselves share
a tie. If two individuals, $i$ and $j$, do not share a tie, then
they can have no effect on one another's disease status that is not
through their contacts' disease statuses. Because $t-b$ is the latest
time at which a disease transmission could affect $Y_{i}^{t}$, we
do not need to condition on the contacts' outcomes past that time.
This assumption implies that the total number of vaccinated and unvaccinated
contacts of individual $i$ who have been sick by day $t-b$ are a
sufficient summary measure of the complete history of all of $i$'s
contacts. It could easily be modified so that the probability of being
infected at any given time depends on a different summary measure,
for example on the proportion of alters who were infectious at or
before time $t-b$.

We also assume that 
\begin{equation}
Y_{i}^{t}\perp V_{j}\mid\left\{ \sum_{m\in\mathcal{A}_{i}:V_{m}=v}Y_{m}^{t-b},\, v=0,1\right\} \,\mbox{for all }j\notin\mathcal{A}_{i}\label{info barrier vac}
\end{equation}
and that, for any covariate $C$ that is required for (\ref{eq:med1})
through (\ref{med4}) to hold, 
\begin{equation}
Y_{i}^{t}\perp C_{j}\mid\left\{ \sum_{m\in\mathcal{A}_{i}:V_{m}=v}Y_{m}^{t-b},\, v=0,1\right\} \,\mbox{for all }j\notin\mathcal{A}_{i}.\label{info barrier cov}
\end{equation}
These assumptions state that any effect of the covariates (including
vaccination) of nodes without ties to $i$ on $i$'s disease status
would again have to be mediated by the disease statuses of $i$'s
contacts. Assumption (\ref{info barrier vac}) implies that the infectiousness
effect is not transitive: whether individual $j$ caught the flu from
a vaccinated or unvaccinated person has no influence on whether individual
$j$ transmits the flu.

Embedded in assumptions (\ref{info barrier})-(\ref{info barrier cov})
is the assumption that all ties are equivalent and all non-ties are
equivalent with respect to transmission of the outcome. This is likely
to be a simplification of reality. It can be relaxed (see Section
5.3), but we make it now for heuristic purposes. It rules out the
possibility that some types of ties, like roommates, are more likely
to facilitate disease transmission than others, like friends who live
in different dorms. It allows an individual to come into contact with
and possibly infect (or be infected by) people with whom he does not
share a tie, but it entails that he will come into contact with any
individual in the network who is not his contact with equal probability.
This rules out, for example, the possibility that an individual is
more likely to be infected by the friends of his friends than by a
distant node on the network. 

We also make the no-unmeasured-confounding assumption that, if there
exists a person with whom two individuals in the network interact
regularly, then that person is also in the network (with ties to both
individuals). In some settings it may be possible to satisfy this
condition, e.g. in full sociometric studies conducted de novo, or
in studies of online data.

\subsection{Estimation and hypothesis testing}

Consider the following strategy for estimating a new contagion and
new infectiousness effect, defined below:
\begin{enumerate}
\item Randomly select from the network $K$ pairs of nodes such that the
two nodes in each pair share a tie, but, for each pair, neither node
nor any of their contacts has a tie to a node in any other pair or
to the contacts of any member of any other pair. The number of possible
such pairs will depend on the network size and topology. In the next
section, we discuss methods for sampling these pairs. Randomly select
one member of each pair to be the ego and one to be the alter. 
\item Index the pairs by $k$, and let $e_{k}$ index the ego and $a_{k}$
the alter in the $k^{th}$ pair. For the $k^{th}$ pair, define a
group, also indexed by $k$, that includes nodes $a_{k}$, $e_{k}$,
$\mathcal{A}_{e_{k}}$, and $\mathcal{A}_{a_{k}}$. That is, it includes
the alter-ego pair and all nodes with ties to either the alter or
the ego. Due to the way we selected pairs, none of the members of
group $k$ can belong to any other group. Below, we suppress the index
$k$ when context allows. As in the sections above, $T_{k}$ is the
time of the first infection in the pair $(a_{k},e_{k})$. Let $ $$\boldsymbol{C}_{k}$
be a collection of covariates for group $k$, where the variables
included in $\boldsymbol{C}$ are precisely those required for assumptions
(\ref{eq:med1}) through (\ref{med4}) to hold for outcome $Y_{e}^{T(1)+b}$,
mediator $Y_{a}^{T(1)}$, and treatment $V_{a}$. Note that $V_{e}$
should be included in $\boldsymbol{C}$ as it is likely to be a confounder
of the mediator - outcome relationship. The number of mutual contacts
of the alter and ego who were sick by time $T-b$ must also be included. 
\item Let $U_{e_{k}}^{T_{k}+f}$ and $L_{e_{k}}^{T_{k}+f}$ be the number
of unvaccinated and vaccinated nodes, respectively, with ties to $e_{k}$
who were sick by time $T_{k}+f$. Define $U_{a_{k}}^{T-b}$ and $L_{a_{k}}^{T-b}$
similarly as the number of unvaccinated and vaccinated nodes, respectively,
with ties to $a_{k}$ who were sick by time $T_{k}-b$. Recall that
$f$ is the infectiousness period and $b$ the incubation period,
defined in Section 3.2.1.
\item Estimate an average modified contagion effect 
\begin{eqnarray*}
Con^{*}=\frac{E\left[Y_{e}^{T(1)+s}(0,Y_{a}^{T(1)}(1))\mid U_{a}^{T-b},L_{a}^{T-b},U_{e}^{T+f},L_{e}^{T+f},\boldsymbol{C}\right]}{E\left[Y_{e}^{T(0)+b}(0,Y_{a}^{T(0)}(0))\mid U_{a}^{T-b},L_{a}^{T-b},U_{e}^{T+f},L_{e}^{T+f},\boldsymbol{C}\right]}
\end{eqnarray*}
and an average modified infectiousness effect 
\begin{eqnarray*}
Inf^{*}=\frac{E\left[Y_{e}^{T(1)+s}(1,Y_{a}^{T(1)}(1))\mid U_{a}^{T-b},L_{a}^{T-b},U_{e}^{T+f},L_{e}^{T+f},\boldsymbol{C}\right]}{E\left[Y_{e}^{T(1)+s}(0,Y_{a}^{T(1)}(1))\mid U_{a}^{T-b},L_{a}^{T-b},U_{e}^{T+f},L_{e}^{T+f},\boldsymbol{C}\right]}
\end{eqnarray*}
and their standard errors. 
\end{enumerate}
Through Step 2, the procedure we described is nearly identical to
the proposal in Section 4, the only difference being that groups are
extracted from a network in Step 1 rather than being independently
ascertained. Consideration for this sampling scheme becomes crucial
when we estimate the parameters of GLMs like (\ref{model1}) and (\ref{model2}).
The standard errors derived from these GLMs are consistent only if
the residuals across groups are uncorrelated. The residuals are indeed
uncorrelated for independent groups, but, in the network setting,
they generally are not. However, the set of additional covariates
introduced in Step 3 essentially blocks the flow of information between
groups. Conditional on these additional covariates, the residuals
are uncorrelated, even in the network setting (see next section for
proof). Roughly, because $U_{e_{k}}^{T_{k}+f}$ and $L_{e_{k}}^{T_{k}+f}$
summarize the disease statuses of the ego's contacts $b$ days before
the outcome \textbf{$Y_{e_{k}}^{T(1)+s}$ }is assessed, conditioning
on them ensures that the outcomes are uncorrelated across groups.
Because\textbf{ }$U_{a_{k}}^{T-b}$ and $L_{a_{k}}^{T-b}$ summarize
the disease statuses of the alter's contacts $b$ days before the
mediator\textbf{ $Y_{a_{k}}^{T(1)}$} is assessed,\textbf{ }conditioning
on them ensures that mediators are uncorrelated across groups. 

The effects defined in Step 4 differ from $Con$ and $Inf$ only in
the conditioning set, but this changes slightly the causal effect
being estimated. Conditioning on $U_{a}^{T-b}$ and $L_{a}^{T-b}$
is just like conditioning on an extra pair of confounders: these variables
occur before the mediator and are independent of the treatment; therefore
they can be considered to be pre-treatment covariates. On the other
hand, $U_{e}^{T+f}$ and $L_{e}^{T+f}$ occur after the mediator and
lie on a possible pathway from the mediator to the outcome. Conditioning
on these variables has the effect of biasing $Con^{*}$ and $Inf^{*}$
towards the null relative to $Con$ and $Inf$, because it blocks
the path from $Y_{a}^{T}$ to $Y_{e}^{T+s}$ that operates when the
alter infects a friend of the ego, who then infects the ego. However,
conditioning on these variables leaves the direct path from $Y_{a}^{T}$
to $Y_{e}^{T+s}$ open, and this path operates whenever the alter
infects the ego directly. Therefore, whenever $Con$ and $Inf$ are
non-null so are $Inf^{*}$ and $Con^{*}$. Hypothesis tests using
$Con^{*}$ and $Inf^{*}$ are conservative and consistent for hypothesis
tests for $Con$ and $Inf$. Similarly, tests that $Con^{*}$ and
$Inf^{*}$ are less than the null value or are greater than the null
value are also valid and consistent for the analogous tests for $Con$
and $Inf$, respectively.

\subsubsection{Justification for the use of GLMs}

Suppose that the models 

\begin{eqnarray}
 &  & g(E\left[Y_{e_{k}}^{T_{k}+s}\mid V_{a_{k}},Y_{a_{k}}^{T_{k}}=1,U_{a}^{T-b},L_{a}^{T-b},U_{e}^{T+f},L_{e}^{T+f},\boldsymbol{C}_{k}\right])\label{M1}\\
 &  & =\beta_{0}+\beta_{1}V_{a_{k}}+\beta_{2}U_{a}^{T-b}+\beta_{3}L_{a}^{T-b}+\beta_{4}U_{e}^{T_{k}+f}+\beta_{5}L_{e}^{T_{k}+f}+\beta_{6}'\boldsymbol{C}_{k}\nonumber 
\end{eqnarray}
 and
\begin{eqnarray}
 &  & m(E\left[Y_{a_{k}}^{T_{k}}\mid V_{a_{k}},U_{a}^{T-b},L_{a}^{T-b},U_{e}^{T+f},L_{e}^{T+f},\boldsymbol{C}_{k}\right])\nonumber \\
 &  & =\alpha_{0}+\alpha_{1}V_{a_{k}}+\alpha_{2}U_{a}^{T_{k}-b}+\alpha_{3}L_{a}^{T_{k}-b}+\alpha_{4}U_{e}^{T_{k}+f}+\alpha_{5}L_{e}^{T_{k}+f}+\alpha_{6}'\boldsymbol{C}_{k}\label{M2}
\end{eqnarray}
are correctly specified for $g()$, $m()$ known link functions. For
the effect on the ratio scale with a binary common outcome like the
flu we would specify $g()$ to be the log link and $m()$ the logit
link, like we did in Sections 3 and 4. We have only to prove that
the residuals from model (\ref{M1}) are uncorrelated with one another
and that the residuals from model (\ref{M2}) are uncorrelated with
one another \citep{breslow1996generalized,gill2001generalized}.
\begin{description}
\item [{Result~1}] Let $Res_{a_{k}}=Y_{a_{k}}^{T_{k}}-m^{-1}\left(\alpha_{0}+\alpha_{1}V_{a_{k}}+\alpha_{2}U_{a}^{T_{k}-b}+\alpha_{3}L_{a}^{T_{k}-b}+\alpha_{4}U_{e}^{T_{k}+f}+\alpha_{5}L_{e}^{T_{k}+f}+\alpha_{6}'\boldsymbol{C}_{k}\right)$.
Then $Res_{a_{k}}$ and $Res_{a_{h}}$ are uncorrelated.
\item [{Proof}] $ $Without loss of generality assume that $T_{k}>T_{h}$.
Under correct specification of (\ref{M2}), $E[Res_{a_{k}}]=E[Res_{a_{h}}]=0$.
Therefore $Cov(Res_{a_{k}},Res_{a_{h}})=E\left[Res_{a_{k}}Res_{a_{h}}\right]$.
Letting $S_{k}$ denote the set of variables $\left\{ V_{a_{k}},U_{a}^{T-b},L_{a}^{T-b},U_{e}^{T+f},L_{e}^{T+f},\boldsymbol{C}_{k}\right\} $,
we have 
\begin{eqnarray*}
 &  & E\left[Res_{a_{k}}Res_{a_{h}}\right]\\
 & = & E\left[E\left[Res_{a_{k}}Res_{a_{h}}\mid S_{k},S_{h}\right]\right]\\
 & = & E\left[E\left[\left\{ Y_{a_{k}}^{T_{k}}-E\left[Y_{a_{k}}^{T_{k}}\mid S_{k}\right]\right\} \left\{ Y_{a_{h}}^{T_{h}}-E\left[Y_{a_{h}}^{T_{h}}\mid S_{h}\right]\right\} \mid S_{k},S_{h}\right]\right]\\
 & = & E\left[E\left[Y_{a_{k}}^{T_{k}}-E\left[Y_{a_{k}}^{T_{k}}\mid S_{k}\right]\mid S_{k},S_{h}\right]\times E\left[Y_{a_{h}}^{T_{h}}-E\left[Y_{a_{h}}^{T_{h}}\mid S_{h}\right]\mid S_{k},S_{h}\right]\right]\\
 & = & E\left[\left\{ E\left[Y_{a_{k}}^{T_{k}}\mid S_{k},S_{h}\right]-E\left[Y_{a_{k}}^{T_{k}}\mid S_{k}\right]\right\} \times E\left\{ Y_{a_{h}}^{T_{h}}-E\left[Y_{a_{h}}^{T_{h}}\mid S_{h}\right]\mid S_{k},S_{h}\right\} \right]\\
 & = & E\left[\left\{ E\left[Y_{a_{k}}^{T_{k}}\mid S_{k}\right]-E\left[Y_{a_{k}}^{T_{k}}\mid S_{k}\right]\right\} \times E\left\{ Y_{a_{h}}^{T_{h}}-E\left[Y_{a_{h}}^{T_{h}}\mid S_{h}\right]\mid S_{k},S_{h}\right\} \right]\\
 & = & 0.
\end{eqnarray*}
The second equality follows from the correct specification of (\ref{M2}).
The third equality holds because, by assumptions (\ref{info barrier}),
(\ref{info barrier vac}), and (\ref{info barrier cov}), $Y_{a_{k}}^{T_{k}}\perp Y_{a_{h}}^{T_{h}}\mid S_{k},S_{h}$.
The fifth inequality holds because $Y_{a_{k}}^{T_{k}}\perp S_{h}\mid S_{k}$,
again by assumptions (\ref{info barrier}), (\ref{info barrier vac}),
and (\ref{info barrier cov}).
\item [{Result~2}] Let $Res_{e_{k}}=Y_{e_{k}}^{T_{k}+s}-g^{-1}\left(\beta_{0}+\beta_{1}V_{a_{k}}+\beta_{2}U_{a_{k}}^{T_{k}-b}+\beta_{3}L_{a_{k}}^{T_{k}-b}+\beta_{4}U_{e}^{T_{k}+f}+\beta_{5}L_{e_{k}}^{T_{k}+f}+\beta_{6}'\boldsymbol{C}_{k}\right)$.
Then $Res_{a_{k}}$ and $Res_{a_{h}}$ are uncorrelated.
\end{description}
The proof of Result 2 is very similar to the proof of Result 1 and
we therefore omit it. It relies on the fact that, conditional on the
fact that $T+f=T+s-b$ and therefore conditioning on $U_{e_{k}}^{T_{k}+f}$$ $
and $L_{e_{k}}^{T_{k}+f}$ satisfies the conditions of assumptions
(\ref{info barrier}), (\ref{info barrier vac}), and (\ref{info barrier cov})
and renders $Y_{e_{k}}^{T_{k}+s}$ independent of outcomes, vaccines,
and covariates for other groups.

\subsubsection{Implementation}

Step 1 is the most difficult to implement. One could enumerate all
possible ways of partitioning the network into non-overlapping groups
comprised of a pair of nodes and all of their contacts, associate
the partitions with a discrete uniform distribution, and randomly
sample one realization of the uniform distribution. Steps 2 and 3
of the testing procedure are perfunctory. If we define $\boldsymbol{C}^{*}=\left(U_{a}^{T-b},L_{a}^{T-b},U_{e}^{T+f},L_{e}^{T+f},\boldsymbol{C}\right)$
to be a new collection of covariates then step 4 proceeds as in Sections
3 and 4. Interactions between components of $\boldsymbol{C}^{*}$
and the other predictors in the model can easily be accommodated.
To test the hypotheses that $Con$ and $Inf$ are null, we estimate
95\% confidence intervals for the modified contagion and infectiousness
estimands ($Con^{*}$ and $Inf^{*}$) based on the estimates and standard
errors calculated in Step 4. We reject the hypothesis that $Con$
is null if our confidence interval for the estimand in $Con^{*}$
does not include the null value and we reject the hypothesis that
$Inf$ is null if our confidence interval for the estimand in $Inf^{*}$
does not include the null value.

\subsection{Relaxing some assumptions}

We assumed throughout that vaccination occurs before the start of
follow-up, but this is not necessary for our methods. If vaccination
can occur during follow-up, define $V_{i}^{t}$ to be an indicator
of having been vaccinated by time $t$. Assume that the effect of
vaccination, including any infectiousness effect, is immediate. If
an individual becomes infectious on day $T$, he would have been infected
on day $T-b$. If he was vaccinated by time $T-b$, then the vaccine
would have been in full effect at the time of infection. Then $V_{a}^{T-b}$
can replace $V_{a}$ as the ``treatment'' in the contagion, infectiousness,
and indirect effects. We similarly redefine the summary measures for
vaccinated and unvaccinated contacts of the alter and ego that appear
in assumptions (\ref{info barrier}) through (\ref{info barrier cov})
and that are included in $\boldsymbol{C}$. Include $V_{e}^{T-b}$
in the set of confounders because the mediator occurs at time $T$
and therefore the ego's vaccination status at time $T-b$ suffices
to control for any confounding. 

We assumed throughout that the infectious and incubation periods ($f$
and $b$) are constant across individuals. These assumptions, along
with the assumption that the effect of vaccination is immediate, could
be relaxed if the determinants of time to efficacy of vaccine, length
of infectious period, and length of incubation period were observed
covariates. In this case we could, for example, infer effective time
of vaccination, incubation period, and infectious period for each
individual based on their covariates.

We assumed in Section 5.2 that the probability of disease transmission
between two connected nodes does not depend on the type of tie. This
assumption can be avoided with the addition of several covariates
to models (\ref{M1}) and (\ref{M2}): we would condition on the type
of tie that exists between the alter and the ego, and also include
separate $U$ and $L$ terms for each type of tie. We also assumed
in Section 5.2 that an individual will come into contact with any
individual in the network who is not his contact with equal probability.
This can be relaxed by expanding the $k$ groups we define in Step
1 of the estimation procedure to include nodes within several degrees
of separation from the alter and ego.

\section{Simulations }

\subsection{Independent groups}

We ran simulations for three different sample sizes, $K=200$, $K=500$,
and $K=1000$ independent groups. Each group comprised an alter, an
ego, and $n_{k}$ mutual contacts. First we generated $K$ contact
group sizes $n_{k}$ by sampling from a Poisson distribution with
mean $\lambda=3$. Next, we assigned vaccination statuses to each
individual in each group, including the alters and egos, with probability
$0.4$. We simulated the behavior of each group during a flu epidemic
over 100 days. For the purposes of the simulation, we assumed that
each member of a group had contact with all other members of the same
group. Each day, an uninfected member of a group had a baseline probability
of $p_{o}$ of being infected from outside of the group, a baseline
probability of $p_{u}$ of being infected by any infectious, unvaccinated
member of the same group and a baseline probability of $p_{v}$ of
being infected by any infectious, vaccinated member of the same group.
If vaccinated, an individual's probability of being infected by any
source was multiplied by $\delta\leq1$. If infected on day $t$,
an individual was infectious from day $t+1$ through day $t+4$ and
incapable of being infected or transmitting infection from day $t+5$
until the end of follow-up. This corresponds to an incubation period
of $b=1$ and an infectious period of $f=3$, and it mimics the flu,
for which the incubation period is between one and three days and
the infectious period is between three and six days \citep{earn2002ecology}.

In all simulations, we fixed $p_{o}=0.01$. We specified two different
simulation settings for the parameters $ $$\delta$, $p_{v}$, and
$p_{u}$, one setting corresponding to the null of no infectiousness
or contagion effects ($ $$\delta=1$; $p_{v}=$$p_{u}=0.4$) and
one setting corresponding to the presence of protective contagion
and infectiousness effects ($ $$\delta=0.1$; $p_{v}=0.5$, $p_{u}=0.05$).
We simulated 500 epidemics each under of the two scenarios, and for
each simulation we estimated the infectiousness and contagion effects
as follows: Among the subset of groups with $Y_{a}^{T}=1$ and using
a log-linear link function, we regressed $Y_{e}^{T+s}$ on $V_{a}$
and on the set of potential confounders comprised by the ego's vaccination
status, the sum $U_{a}^{T-b}$ of unvaccinated mutual contacts who
were infectious at time $T-b$, and the sum $L_{a}^{T-b}$ of vaccinated
mutual contacts who were infectious at time $T-b$. We regressed $Y_{a}^{T}$
on the same covariates using a logistic link function. The contagion
and infectiousness effects are identified by the expressions given
in $ $(\ref{con params}) and (\ref{inf params}), evaluated at the
sample mean value of the covariates $U_{a}^{T-b}$ and $L_{a}^{T-b}$.
We bootstrapped the standard errors with 500 bootstrap replications.

\begin{table}

\caption{Simulation results for independent groups}

\medskip{}

\medskip{}

\begin{centering}
\begin{tabular}{|c|c|c|c|c|}
\multicolumn{5}{c}{Under $H_{0}$}\tabularnewline
\hline 
Number of groups & Infectiousness (SE) & Coverage & Contagion (SE) & Coverage\tabularnewline
\hline 
\hline 
\multirow{1}{*}{$K=200$} & 1.014 (0.138) & 94\% & 1.016 (0.202) & 94\%\tabularnewline
\hline 
\multirow{1}{*}{$K=500$} & 1.001 (0.082) & 92.2\% & 0.997 (0.117) & 93.6\%\tabularnewline
\hline 
\multirow{1}{*}{$K=1000$} & 0.997 (0.057) & 95\% & 1.001 (0.083) & 94\%\tabularnewline
\hline 
\end{tabular}
\par\end{centering}

\medskip{}

\medskip{}

\centering{}%
\begin{tabular}{|c|c|c|c|c|}
\multicolumn{5}{c}{Under $H_{A}$}\tabularnewline
\hline 
Number of groups & Infectiousness (SE) & Power ~~~ & Contagion (SE) & Power~~~~~\tabularnewline
\hline 
\hline 
\multirow{1}{*}{$K=200$} & 0.453 (1.160) & 49\% & 0.258 (0.079) & 100\%\tabularnewline
\hline 
\multirow{1}{*}{$K=500$} & 0.443 (0.154) & 86\% & 0.255 (0.049) & 100\%\tabularnewline
\hline 
\multirow{1}{*}{$K=1000$} & 0.445 (0.107) & 100\% & 0.258 (0.034) & 100\%\tabularnewline
\hline 
\end{tabular}
\end{table}

The results are given in Table 1. For each simulation setting, that
is, for each sample size ($K$) and for both the null hypothesis and
the alternative hypothesis, we present the mean point estimates for
the infectiousness and contagion effects on the ratio scale, the mean
bootstrap standard error estimator, and the percent coverage of the
95\% confidence interval based on the $2.5^{th}$ and $97.5^{th}$
bootstrap quantiles. For simulations under the null hypothesis, we
report coverage and for simulations under the alternative we report
power, given by $100\%$ minus the percent coverage. The point estimates
are stable across sample sizes and the coverage of the basic bootstrap
confidence interval is close to 95\% under the null for all $K$.
The power under the alternative is $100\%$ for the contagion effect,
but for the infectiousness effect power is low (49\%) when $K=200$.

\subsection{Social network data}

The procedure proposed in Section 5.2 for hypothesis testing using
social network data suffers from low power. In part this is because
$Con^{*}$ and $Inf^{*}$ are biased towards the null relative to
$Con$ and $Inf$, but the primary reason for the loss of power is
the extraction of conditionally independent pairs of nodes from the
network. As the simulation illustrates, this results in a dramatic
reduction in the sample size used for analysis. Because infectious
outcomes sampled from nodes in a network are dependent, the effective
sample size for inference about such outcomes will always be smaller
than the observed number of nodes, and how much more information about
the parameters of interest is available depends on the specific setting.
Important areas for future research include determining the effective
sample size when observations are sampled from a network and are therefore
dependent, and developing methods that make use of all available information. 

We ran simulations for three different network sizes: $12000$ nodes,
$10000$ nodes, and $8000$ nodes. We simulated a network of $10000$
nodes as follows: first, we simulated $2000$ independent groups of
$5$ nodes, with each group being fully connected (i.e. there are
ties between each pair of nodes in the group of 5). For each node
we then added a tie to each out-of-group node with probability $0.0001$.
Because ties are undirected (if node $i$ is tied to node $j$, then
by definition node $j$ is tied node $i$), this results in approximately
2 expected out-of-group ties per node. To simulate networks of size
$12000$ and $8000$, we simulated $2400$ and $1600$ independent
groups, respectively, and scaled the probability of an out-of-group
tie to maintain an expected value of approximately $2$ for each node.
This network structure could represent a sample of families living
in a city, where individuals are fully connected to the members of
their family and occasionally connected to members of other families.
After running step 1 of the procedure outlined in Section 5.2, we
were left with $K=707$ alter-ego pairs for the network of size $12000$,
$K=581$ for the network of size $10000$, and $K=466$ for the network
of size $8000$. 

On each of these three fixed networks, we simulated 200 epidemics
under the null of no infectiousness or contagion effect and $200$
epidemics under the alternative. For each simulation, we assigned
vaccination statuses to each individual in the network with probability
$0.5$. We then simulated the behavior of each group during a flu
epidemic over 100 days. An uninfected node had a probability of $p_{o}=0.01$
of being infected from outside of the network on day 1 and there were
no outside infections thereafter. Under the alternative, on each day
an uninfected node had a baseline probability of $p_{u}=0.5$ of being
infected by any infectious, unvaccinated contact and group and a baseline
probability of $p_{v}=0.01$ of being infected by any infectious,
vaccinated contact. If vaccinated, an individual's probability of
being infected by any source was multiplied by $\delta=0.2$. Under
the null, on each day an uninfected node had a probability of $p_{u}=p_{v}=0.5$
of being infected by any infectious contact (that is, node with which
it shared a tie). To ensure that the contagion effect was null, we
specified that $\delta=1$, that is, that vaccination had no protective
effect against contracting the flu. In both settings, if infected
on day $t$ an individual was infectious from day $t+1$ through day
$t+4$ and incapable of being infected or transmitting infection from
day $t+5$ until the end of follow-up. 

For each simulation, we estimated the infectiousness and contagion
effects following the procedure described in Section 5.2. We evaluated
these effects at the sample mean value of the covariates $U_{a}^{T-b}$,
$L_{a}^{T-b}$, $U_{e}^{T+f}$ and $L_{e}^{T+f}$. We bootstrapped
the standard errors with 1000 bootstrap replications. The results
are given in Table 2. For each simulation setting, that is for each
network size and for both the null hypothesis and the alternative
hypothesis, we present the mean point estimates for the infectiousness
and contagion effects on the ratio scale, the mean bootstrap standard
error estimator, and the percent coverage of the 95\% confidence interval
based on the $2.5^{th}$ and $97.5^{th}$ bootstrap quantiles. For
simulations under the alternative hypothesis we calculated the power,
given by $100\%$ minus the percent coverage. For the $8000$- and
$10000$-node networks, there were 6 and 1 simulations, respectively,
out of 200, for which the GLMs used to estimate the parameters involved
in the contagion and infectiousness effects did not converge due to
empty strata of the predictors. We omit these simulations from the
results in Table 2, but note that in a extreme cases convergence could
be an issue in addition to power. 

\begin{table}
\caption{Simulation results for network data}

\begin{centering}
\medskip{}

\par\end{centering}

\medskip{}

\begin{centering}
\begin{tabular}{|c|c|c|c|c|}
\multicolumn{5}{c}{Under $H_{0}$}\tabularnewline
\hline 
Network size & Infectiousness (SE) & Coverage & Contagion (SE) & Coverage\tabularnewline
\hline 
\hline 
\multirow{1}{*}{$8000$ nodes} & 0.996 (0.001) & 100\% & 1.205 (1.657) & 96\%\tabularnewline
\hline 
\multirow{1}{*}{$10000$ nodes} & 1.000 (0.001) & 100\% & 1.183 (1.183) & 94\%\tabularnewline
\hline 
\multirow{1}{*}{$12000$ nodes} & 1.001 (0.001)  & 100\% & 1.166 (0.) & 94\%\tabularnewline
\hline 
\end{tabular}
\par\end{centering}

\begin{centering}
\medskip{}

\par\end{centering}

\begin{centering}
\medskip{}

\par\end{centering}

\centering{}%
\begin{tabular}{|c|c|c|c|c|}
\multicolumn{5}{c}{Under $H_{A}$}\tabularnewline
\hline 
Network size & Infectiousness (SE) & Power~~~~ & Contagion (SE) & Power~~~~~\tabularnewline
\hline 
\hline 
\multirow{1}{*}{$8000$ nodes} & 0.650 (0.259) & 45\% & 0.168 (0.017) & 99\%\tabularnewline
\hline 
\multirow{1}{*}{$10000$ nodes} & 0.616 (0.072) & 53\% & 0.164 (0.013) & 100\%\tabularnewline
\hline 
\multirow{1}{*}{$12000$ nodes} & 0.609 (0.054) & 63\% & 0.164 (0.010) & 100\%\tabularnewline
\hline 
\end{tabular}
\end{table}

The point estimates are stable across network sizes and the coverage
of the basic bootstrap confidence interval is close to or above 95\%
under the null for all network sizes. The power under the alternative
is close to $100\%$ for the contagion effect for all network sizes,
but for the infectiousness effect power is low: $45\%$ for the network
of size $8000$, increasing to $63\%$ for the network of size $12000$.

One concern that has been raised about previous uses of statistical
models like GLMs and GEEs for network data is the possibility that
the models lack any power to reject the null hypothesis when the alternative
is true \citep{shalizi2012statistics}. This is a concern because
the models are inherently misspecified under the alternative hypothesis,
even if they are correctly specified under the null hypothesis. Because
the methods we propose here can be correctly specified under both
the null and the alternative hypotheses, they can be powered to reject
the null hypothesis when the infectiousness or contagion effect is
present.

\section{Discussion}

We proposed methods for consistently estimating contagion and infectiousness
effects in independent groups of arbitrary size; these methods are
easy to implement and perform well in simulations. We extended our
methodology to groups sampled from social network data, providing
a theoretically justified method for using GLMs to analyze network
data. Note that the principles we applied to GLMs can be applied to
GEEs as well, resulting in correctly specified GEEs for network data.
The principles that justify our use of GLMs to estimate the contagion
and infectiousness effects are easily extended to any estimand for
which GLMs would be a desirable modeling tool. However, our network
data methods require a large amount of data and are not appropriate
for small or dense networks. On the one hand this highlights the fact
that dependence among observations in networks reduces effective sample
size and necessitates larger samples; on the other hand methods should
be developed that can harness more information from the data and increase
the power to detect contagion, infectiousness, and other causal effects.

\bibliographystyle{jasa}
\bibliography{references}

\begin{thebibliography}{47}
\newcommand{\enquote}[1]{``#1''}
\expandafter\ifx\csname natexlab\endcsname\relax\def\natexlab#1{#1}\fi
\expandafter\ifx\csname url\endcsname\relax
  \def\url#1{{\tt #1}}\fi
\expandafter\ifx\csname urlprefix\endcsname\relax\def\urlprefix{URL }\fi

\bibitem[{Airoldi et~al.(2013)Airoldi, Toulis, Kao, and
  Rubin}]{airoldi2013estimation}
Airoldi, E., Toulis, P., Kao, E., and Rubin, D.~B.
\newblock \enquote{Estimation of causal peer influence effects.}
\newblock In {\em Proceedings of the 30th International Conference on Machine
  Learning, Atlanta, GA, JMLR: W\&CP\/}, volume~28 (2013).

\bibitem[{Ali and Dwyer(2009)}]{ali2009estimating}
Ali, M.~M. and Dwyer, D.~S.
\newblock \enquote{Estimating peer effects in adolescent smoking behavior: A
  longitudinal analysis.}
\newblock {\em Journal of Adolescent Health\/}, 45(4):402--408 (2009).

\bibitem[{Anderson et~al.(1985)Anderson, May et~al.}]{anderson1985vaccination}
Anderson, R.~M., May, R.~M., et~al.
\newblock \enquote{Vaccination and herd immunity to infectious diseases.}
\newblock {\em Nature\/}, 318(6044):323--329 (1985).

\bibitem[{Aronow and Samii(2012)}]{aronow2012estimating}
Aronow, P.~M. and Samii, C.
\newblock \enquote{Estimating average causal effects under general
  interference.}
\newblock In {\em Summer Meeting of the Society for Political Methodology,
  University of North Carolina, Chapel Hill, July\/}, 19--21. Citeseer (2012).

\bibitem[{Bowers et~al.(2013)Bowers, Fredrickson, and
  Panagopoulos}]{bowers2013reasoning}
Bowers, J., Fredrickson, M.~M., and Panagopoulos, C.
\newblock \enquote{Reasoning about Interference Between Units: A General
  Framework.}
\newblock {\em Political Analysis\/}, 21(1):97--124 (2013).

\bibitem[{Breslow(1996)}]{breslow1996generalized}
Breslow, N.~E.
\newblock \enquote{Generalized linear models: checking assumptions and
  strengthening conclusions.}
\newblock {\em Statistica Applicata\/}, 8:23--41 (1996).

\bibitem[{Cacioppo et~al.(2009)Cacioppo, Fowler, and
  Christakis}]{cacioppo2009alone}
Cacioppo, J.~T., Fowler, J.~H., and Christakis, N.~A.
\newblock \enquote{Alone in the crowd: the structure and spread of loneliness
  in a large social network.}
\newblock {\em Journal of personality and social psychology\/}, 97(6):977
  (2009).

\bibitem[{Christakis and Fowler(2007)}]{christakis2007spread}
Christakis, N.~A. and Fowler, J.~H.
\newblock \enquote{The spread of obesity in a large social network over 32
  years.}
\newblock {\em New England Journal of Medicine\/}, 357(4):370--379 (2007).

\bibitem[{Christakis and Fowler(2008)}]{christakis2008collective}
---.
\newblock \enquote{The collective dynamics of smoking in a large social
  network.}
\newblock {\em New England journal of medicine\/}, 358(21):2249--2258 (2008).

\bibitem[{Christakis and Fowler(2013)}]{christakis2013social}
---.
\newblock \enquote{Social contagion theory: examining dynamic social networks
  and human behavior.}
\newblock {\em Statistics in Medicine\/}, 32(4):556--577 (2013).

\bibitem[{Cohen-Cole and Fletcher(2008)}]{cohen2008obesity}
Cohen-Cole, E. and Fletcher, J.~M.
\newblock \enquote{Is obesity contagious? Social networks vs. environmental
  factors in the obesity epidemic.}
\newblock {\em Journal of Health Economics\/}, 27(5):1382--1387 (2008).

\bibitem[{Eames and Keeling(2002)}]{eames2002modeling}
Eames, K. T.~D. and Keeling, M.~J.
\newblock \enquote{Modeling dynamic and network heterogeneities in the spread
  of sexually transmitted diseases.}
\newblock {\em Proceedings of the National Academy of Sciences\/},
  99(20):13330--13335 (2002).

\bibitem[{Eames and Keeling(2004)}]{eames2004monogamous}
---.
\newblock \enquote{Monogamous networks and the spread of sexually transmitted
  diseases.}
\newblock {\em Mathematical Biosciences\/}, 189(2):115--130 (2004).

\bibitem[{Earn et~al.(2002)Earn, Dushoff, and Levin}]{earn2002ecology}
Earn, D. J.~D., Dushoff, J., and Levin, S.~A.
\newblock \enquote{Ecology and evolution of the flu.}
\newblock {\em Trends in Ecology \& Evolution\/}, 17(7):334--340 (2002).

\bibitem[{Eubank et~al.(2004)Eubank, Guclu, Kumar, Marathe, Srinivasan,
  Toroczkai, and Wang}]{eubank2004modelling}
Eubank, S., Guclu, H., Kumar, V. S.~A., Marathe, M.~V., Srinivasan, A.,
  Toroczkai, Z., and Wang, N.
\newblock \enquote{Modelling disease outbreaks in realistic urban social
  networks.}
\newblock {\em Nature\/}, 429(6988):180--184 (2004).

\bibitem[{Fine(1993)}]{fine1993herd}
Fine, P. E.~M.
\newblock \enquote{Herd immunity: history, theory, practice.}
\newblock {\em Epidemiologic Reviews\/}, 15(2):265--302 (1993).

\bibitem[{Fowler and Christakis(2008)}]{fowler2008estimating}
Fowler, J.~H. and Christakis, N.~A.
\newblock \enquote{Estimating peer effects on health in social networks: A
  response to Cohen-Cole and Fletcher; Trogdon, Nonnemaker, Pais.}
\newblock {\em Journal of health economics\/}, 27(5):1400 (2008).

\bibitem[{Gill(2001)}]{gill2001generalized}
Gill, J.
\newblock {\em Generalized Linear Models: A Unified Approach\/}, volume 134.
\newblock Sage Publications, Inc (2001).

\bibitem[{Halloran and Hudgens(2012)}]{halloran2011causal}
Halloran, M.~E. and Hudgens, M.~G.
\newblock \enquote{Causal inference for vaccine effects on infectiousness.}
\newblock {\em International Journal of Biostatistics\/}, 8(2) (2012).

\bibitem[{Halloran and Struchiner(1991)}]{halloran1991study}
Halloran, M.~E. and Struchiner, C.~J.
\newblock \enquote{Study designs for dependent happenings.}
\newblock {\em Epidemiology\/}, 2(5):331--338 (1991).

\bibitem[{Halloran and Struchiner(1992)}]{halloran1992modeling}
---.
\newblock \enquote{Modeling transmission dynamics of stage-specific malaria
  vaccines.}
\newblock {\em Parasitology Today\/}, 8(3):77--85 (1992).

\bibitem[{Halloran and Struchiner(1995)}]{halloran1995causal}
---.
\newblock \enquote{Causal inference in infectious diseases.}
\newblock {\em Epidemiology\/}, 6(2):142--151 (1995).

\bibitem[{Imai et~al.(2010)Imai, Keele, and Tingley}]{imai2010general}
Imai, K., Keele, L., and Tingley, D.
\newblock \enquote{A general approach to causal mediation analysis.}
\newblock {\em Psychological Methods\/}, 15(4):309--334 (2010).

\bibitem[{John and Samuel(2000)}]{john2000herd}
John, T.~J. and Samuel, R.
\newblock \enquote{Herd immunity and herd effect: new insights and
  definitions.}
\newblock {\em European Journal of Epidemiology\/}, 16(7):601--606 (2000).

\bibitem[{Keeling and Eames(2005)}]{keeling2005networks}
Keeling, M.~J. and Eames, K.~T.
\newblock \enquote{Networks and epidemic models.}
\newblock {\em Journal of the Royal Society Interface\/}, 2(4):295--307 (2005).

\bibitem[{Keller and Stiehm(2000)}]{keller2000passive}
Keller, M.~A. and Stiehm, E.~R.
\newblock \enquote{Passive immunity in prevention and treatment of infectious
  diseases.}
\newblock {\em Clinical Microbiology Reviews\/}, 13(4):602--614 (2000).

\bibitem[{Klovdahl(1985)}]{klovdahl1985social}
Klovdahl, A.~S.
\newblock \enquote{Social networks and the spread of infectious diseases: the
  AIDS example.}
\newblock {\em Social Science and Medicine\/}, 21(11):1203--1216 (1985).

\bibitem[{Klovdahl et~al.(1994)Klovdahl, Potterat, Woodhouse, Muth, Muth, and
  Darrow}]{klovdahl1994social}
Klovdahl, A.~S., Potterat, J.~J., Woodhouse, D.~E., Muth, J.~B., Muth, S.~Q.,
  and Darrow, W.~W.
\newblock \enquote{Social networks and infectious disease: The Colorado Springs
  study.}
\newblock {\em Social Science and Medicine\/}, 38(1):79--88 (1994).

\bibitem[{Latora et~al.(2006)Latora, Nyamba, Simpore, Sylvette, Diane, Sylvere,
  and Musumeci}]{latora2006network}
Latora, V., Nyamba, A., Simpore, J., Sylvette, B., Diane, S., Sylvere, B., and
  Musumeci, S.
\newblock \enquote{Network of sexual contacts and sexually transmitted HIV
  infection in Burkina Faso.}
\newblock {\em Journal of Medical Virology\/}, 78(6):724--729 (2006).

\bibitem[{Lazer et~al.(2010)Lazer, Rubineau, Chetkovich, Katz, and
  Neblo}]{lazer2010coevolution}
Lazer, D., Rubineau, B., Chetkovich, C., Katz, N., and Neblo, M.
\newblock \enquote{The coevolution of networks and political attitudes.}
\newblock {\em Political Communication\/}, 27(3):248--274 (2010).

\bibitem[{Lyons(2011)}]{lyons2010spread}
Lyons, R.
\newblock \enquote{The spread of evidence-poor medicine via flawed
  social-network analysis.}
\newblock {\em Statistics, Politics, and Policy\/}, 2(1) (2011).

\bibitem[{Noel and Nyhan(2011)}]{noel2011unfriending}
Noel, H. and Nyhan, B.
\newblock \enquote{The unfriending problem: The consequences of homophily in
  friendship retention for causal estimates of social influence.}
\newblock {\em Social Networks\/}, 33(3):211--218 (2011).

\bibitem[{O'Brien and Dagan(2003)}]{o2003potential}
O'Brien, K. and Dagan, R.
\newblock \enquote{The potential indirect effect of conjugate pneumococcal
  vaccines.}
\newblock {\em Vaccine\/}, 21(17-18):1815--1825 (2003).

\bibitem[{Ogburn and VanderWeele(2013)}]{ogburnDAGs}
Ogburn, E.~L. and VanderWeele, T.~J.
\newblock \enquote{Causal diagrams for interference.}
\newblock Technical report (2013).

\bibitem[{Pearl(2001)}]{pearl2001direct}
Pearl, J.
\newblock \enquote{Direct and indirect effects.}
\newblock In {\em Proceedings of the Seventeenth Conference on Uncertainty in
  Artificial Intelligence\/}, 411--420 (2001).

\bibitem[{Robins and Greenland(1992)}]{robins1992identifiability}
Robins, J.~M. and Greenland, S.
\newblock \enquote{Identifiability and exchangeability for direct and indirect
  effects.}
\newblock {\em Epidemiology\/}, 3(2):143--155 (1992).

\bibitem[{Robins and Richardson(2010)}]{RobinsRichardsonBook}
Robins, J.~M. and Richardson, T.~S.
\newblock \enquote{Alternative graphical causal models and the identification
  of direct effects.}
\newblock In Shrout, P. (ed.), {\em Causality and Psychopathology: Finding the
  Determinants of Disorders and Their Cures\/}. Oxford University Press (2010).

\bibitem[{Rosenbaum(2007)}]{rosenbaum2007interference}
Rosenbaum, P.
\newblock \enquote{Interference between units in randomized experiments.}
\newblock {\em Journal of the American Statistical Association\/},
  102(477):191--200 (2007).

\bibitem[{Rosenquist et~al.(2010)Rosenquist, Murabito, Fowler, and
  Christakis}]{rosenquist2010spread}
Rosenquist, J.~N., Murabito, J., Fowler, J.~H., and Christakis, N.~A.
\newblock \enquote{The spread of alcohol consumption behavior in a large social
  network.}
\newblock {\em Annals of Internal Medicine\/}, 152(7):426--433 (2010).

\bibitem[{Shalizi(2012)}]{shalizi2012statistics}
Shalizi, C.~R.
\newblock \enquote{Comment on "Why and When 'Flawed' Social Network Analyses
  Still Yield Valid Tests of no Contagion".}
\newblock {\em Statistics, Politics, and Policy\/}, 3(1) (2012).

\bibitem[{Shalizi and Thomas(2011)}]{shalizi2011homophily}
Shalizi, C.~R. and Thomas, A.~C.
\newblock \enquote{Homophily and contagion are generically confounded in
  observational social network studies.}
\newblock {\em Sociological Methods \& Research\/}, 40(2):211--239 (2011).

\bibitem[{Valeri and VanderWeele(2013)}]{valeri2012}
Valeri, L. and VanderWeele, T.~J.
\newblock \enquote{Mediation analysis allowing for exposure-mediator
  interactions and causal interpretation: theoretical assumptions and
  implementation with SAS and SPSS macros.}
\newblock {\em Psychological Methods\/}, 18(2):137--150 (2013).

\bibitem[{VanderWeele(2011)}]{vanderweele2011sensitivity}
VanderWeele, T.~J.
\newblock \enquote{Sensitivity analysis for contagion effects in social
  networks.}
\newblock {\em Sociological Methods \& Research\/}, 40(2):240--255 (2011).

\bibitem[{VanderWeele et~al.(2012{\natexlab{a}})VanderWeele, Ogburn, and
  Tchetgen~Tchetgen}]{vanderweele2012and}
VanderWeele, T.~J., Ogburn, E.~L., and Tchetgen~Tchetgen, E.~J.
\newblock \enquote{Why and When" Flawed" Social Network Analyses Still Yield
  Valid Tests of no Contagion.}
\newblock {\em Statistics, Politics, and Policy\/}, 3(1):1--11
  (2012{\natexlab{a}}).

\bibitem[{VanderWeele and
  Tchetgen~Tchetgen(2011{\natexlab{a}})}]{vanderweele2011bounding}
VanderWeele, T.~J. and Tchetgen~Tchetgen, E.~J.
\newblock \enquote{Bounding the Infectiousness Effect in Vaccine Trials.}
\newblock {\em Epidemiology\/}, 22(5):686 (2011{\natexlab{a}}).

\bibitem[{VanderWeele and
  Tchetgen~Tchetgen(2011{\natexlab{b}})}]{vanderweele2011effect}
---.
\newblock \enquote{Effect partitioning under interference in two-stage
  randomized vaccine trials.}
\newblock {\em Statistics \& Probability Letters\/}, 81(7):861--869
  (2011{\natexlab{b}}).

\bibitem[{VanderWeele et~al.(2012{\natexlab{b}})VanderWeele, Tchetgen~Tchetgen,
  and Halloran}]{vanderweele2011components}
VanderWeele, T.~J., Tchetgen~Tchetgen, E.~J., and Halloran, M.~E.
\newblock \enquote{Components of the indirect effect in vaccine trials:
  identification of contagion and infectiousness effects.}
\newblock {\em Epidemiology\/}, 23(5):751--761 (2012{\natexlab{b}}).

\end{thebibliography}

\end{document}